\documentclass[11pt]{article}

\usepackage[margin=1in]{geometry}
\usepackage[T1]{fontenc}
\usepackage{lmodern}

\usepackage{amsmath,amssymb,amsthm,mathtools,bm}
\usepackage{graphicx}
\usepackage{subcaption}
\usepackage{caption}
\usepackage{booktabs}
\usepackage{siunitx}

\usepackage{array}
\usepackage{enumitem}
\usepackage{algorithm}
\usepackage{algorithmic}
\usepackage{hyperref}

\hypersetup{colorlinks=true, linkcolor=blue, citecolor=blue, urlcolor=blue}

\title{Curvelet-Regularized SPDE Inversion on Piecewise-Planar Fractures with Trace-Graph Coupling}
\author{J.\,J.\,Segura}
\date{\today}

\theoremstyle{plain}
\newtheorem{theorem}{Theorem}
\newtheorem{proposition}{Proposition}

\theoremstyle{definition}

\theoremstyle{remark}

\newcommand{\R}{\mathbb{R}}

\newcommand{\argmin}{\operatornamewithlimits{arg\,min}}
\newcommand{\diag}{\operatorname{diag}}
\newcommand{\nnz}{\operatorname{nnz}}
\newcommand{\abs}[1]{\left|#1\right|}
\newcommand{\norm}[1]{\left\|#1\right\|}

\begin{document}
\maketitle

\begin{abstract}
We formulate a sparse-to-dense reconstruction layer for fractured media in which sparse point measurements are mapped onto piecewise-planar fracture supports inferred from 3D trace polylines. Each plane is discretized in local coordinates and estimated via a convex objective that combines a grid SPDE/GMRF quadratic prior with an $\ell_1$ penalty on undecimated discrete curvelet coefficients, targeting anisotropic, fracture-aligned structure that is poorly represented by isotropic smoothness alone. We further define an along-fracture distance through trace-network geodesics and express connectivity-driven regularization as a quadratic form $z^\top P^\top L_G P z$, where $L_G$ is a graph Laplacian on the trace network and $P$ maps plane grids to graph nodes; plane intersections are handled by linear consistency constraints sampled along intersection lines. The resulting optimization admits efficient splitting: sparse linear solves for the quadratic block and coefficient-wise shrinkage for the curvelet block, with standard ADMM convergence under convexity. We specify reproducible synthetic benchmarks, baselines, ablations, and sensitivity studies that isolate directional sparsity and connectivity effects, and provide reference code to generate the figures and quantitative tables.

\end{abstract}

\vspace{0.5em}

\section{Introduction and Motivation}

Classical geostatistics often models correlation as a function of Euclidean distance $\norm{\bm{x}-\bm{x}'}$ in $\R^d$,
leading to variogram-based kriging and Gaussian random field (GRF) models \cite{Cressie1993,Wackernagel2003,ChilesDelfiner2012}.
Fractured media violate this assumption: transport, flow, and even geochemical alteration frequently propagate preferentially
\emph{along connected fractures} and their intersections \cite{Berkowitz2002,Neuman2005,NationalResearchCouncil1996}.
Two points close in Euclidean space may be weakly coupled if separated by intact rock, while distant points may be strongly coupled
if connected by a fracture corridor.

This motivates two complementary principles:

\begin{enumerate}[label=(P\arabic*), leftmargin=*]
\item \textbf{Along-fracture distance:} define smoothness and correlation on a \emph{fracture network} (polyline graph) and treat
Euclidean distance as secondary.
\item \textbf{Geometric sparsity:} fracture-controlled fields often consist of curvilinear, multiscale, directional features (ridges,
fronts, channels) that are sparse in directional multiscale frames such as curvelets (and related systems like ridgelets and shearlets)
\cite{CandesDonoho2004,CandesDonoho2000,CandesDemanetDonohoYing2006,CandesDonoho1999Ridgelets,KutyniokLabate2012}.
\end{enumerate}

\paragraph{Scope statement.}
This paper is a \emph{methods framework} with a working Version~1 architecture: per-plane inference from 3D trace polylines and sparse
samples, solved with a Gaussian baseline and an ADMM curvelet-regularized model. We also formalize the Version~2 extension: explicit
along-fracture regularization via graph Laplacians and coupling across plane intersections. We do not overclaim physics completeness:
the framework is a statistically controlled interpolation/inversion layer intended to be coupled later to flow/transport models.

\section{Related Work and Positioning}

\paragraph{Fractured media, DFNs, and connectivity-driven transport.}
Connectivity is a dominant control on fractured-rock transport and effective properties \cite{Berkowitz2002,Neuman2005,deDreuzy2001}.
DFN frameworks and rock-joint system models provide geometric/statistical representations of fracture sets \cite{DershowitzEinstein1988,LaPointeHudson1985,NationalResearchCouncil1996}.
Our focus is different: given observed traces/planes and sparse property samples, reconstruct a fracture-conditioned property field.

\paragraph{SPDE/GMRF baselines.}
Sparse precision formulations and the SPDE approach link GRFs (e.g., Mat\'ern) to GMRFs, enabling scalable inference on grids
\cite{RueHeld2005,LindgrenRueLindstrom2011}. We use this as a baseline and diagnostic reference.

\paragraph{Directional multiscale transforms in inverse problems and geoscience.}
Curvelets are near-optimal for representing piecewise smooth functions with $C^2$ edges \cite{CandesDonoho2004} and are standard
in sparsity-promoting inverse problems \cite{StarckMurtaghFadili2010,Mallat2009,BoydADMM2011}. In geophysics, curvelet frames
are known to sparsify wavefront-like seismic events and support data recovery/interpolation \cite{HennenfentHerrmann2008,GeophysicsCurvelet2008}.

\paragraph{Graphs and network priors.}
Graph Laplacians provide canonical smoothness functionals on networks and support diffusion/spectral methods \cite{Chung1997,Shuman2013}.
This aligns naturally with ``along-fracture distance''.

\section{Data Model}\label{sec:data}

\subsection{Inputs}

\paragraph{3D trace polylines.}
We assume fracture traces are provided as polylines in $\R^3$:
\[
\mathcal{T}=\{ \gamma_k \}_{k=1}^K,\qquad
\gamma_k = (\bm{x}_{k,1}, \bm{x}_{k,2}, \ldots, \bm{x}_{k,N_k}),\quad \bm{x}_{k,i}\in \R^3.
\]
Each $\gamma_k$ is interpreted as a trace of a fracture plane (or a segment of it) or as geometry lying near a plane.

\paragraph{Sparse samples.}
We also have sparse measurements of a scalar property:
\[
\mathcal{S} = \{ (\bm{x}_j, y_j, \sigma_j)\}_{j=1}^N,
\]
with $\bm{x}_j\in\R^3$ a location, $y_j\in\R$ the observed value, and $\sigma_j>0$ an optional noise standard deviation.

\subsection{Physical meaning of the latent field $z_p(u,v)$}
\label{sec:physical}

The quantity estimated by our framework is a \emph{latent scalar field on a fracture surface}.
In Version~1 we work plane-by-plane: for each inferred plane $p$ we define a local chart $(u,v)$ by an orthonormal basis
$\{\bm{e}_{p,1},\bm{e}_{p,2}\}$ and an origin $\bm{x}_{p,0}$ on the plane, so that any 3D point $\bm{x}$ assigned to plane $p$
is mapped by projection to
\begin{equation}
u(\bm{x})=\bm{e}_{p,1}^{\top}(\bm{x}-\bm{x}_{p,0}), \qquad
v(\bm{x})=\bm{e}_{p,2}^{\top}(\bm{x}-\bm{x}_{p,0}).
\end{equation}
The unknown is then $z_p(u,v)$, discretized on a 2D grid. When we write $z(\bm{x})$ informally, we mean the pullback
$z(\bm{x}) := z_{p(\bm{x})}(u(\bm{x}),v(\bm{x}))$ under the current plane assignment $p(\bm{x})$.

\paragraph{What does $z_p$ represent physically?}
Its meaning is determined entirely by the \emph{measurement process} encoded in the data term (Section~\ref{sec:data}): we assume
we observe sparse noisy samples $y_j \approx z(\bm{x}_j)$ (or a known transform thereof), and we regularize the completion of this field.
The method is therefore an interpolation/inversion \emph{layer} that can be coupled to physics, but does not replace a flow/transport solver.

Concrete examples include:
(i) $z_p=\log T$ (log-transmissivity) or $z_p=\log b$ (log-aperture) inferred from hydraulic tests on fractures;
(ii) geochemical grade, alteration intensity, or damage indices measured at sparse locations on mapped fractures;
(iii) a time-\emph{fixed} tracer concentration (snapshot) or a scalar summary of a transient (e.g.\ peak concentration or arrival time).
For genuinely time-dependent processes, one would estimate a sequence $z_p(u,v,t)$ (multiple independent inversions) or embed the prior
into a PDE-constrained inversion; this is beyond Version~1 and is part of the motivation for the ``physics coupling'' roadmap in Section~\ref{sec:limitations}.

\paragraph{Why along-fracture metrics matter.}
The graph-geodesic/Laplacian constructions (Section~\ref{sec:graph}) are appropriate when the \emph{effective correlation} of the target
quantity is governed by connectivity along fractures (e.g.\ hydraulic/thermal transport corridors), rather than by Euclidean proximity in 3D.
If the target quantity is controlled by different physics (e.g.\ isotropic measurement noise on a planar laboratory proxy), Euclidean priors may suffice.
\subsection{Notation summary}
\label{sec:notation}
In per-plane derivations (Sections 6--7) we drop the plane subscript $p$ for readability; all operations are per-plane unless stated otherwise.

\begin{table}[h]
\centering
\caption{Core symbols and dimensions used throughout.}
\label{tab:notation}
\begin{tabular}{@{}>{\raggedright\arraybackslash}p{0.16\textwidth}p{0.74\textwidth}@{}}
\toprule
Symbol & Meaning \\ \midrule
$p$ & plane index; each fracture plane (or plane cluster) is solved on its own grid in Version~1 \\
$(u,v)$ & local coordinates on a plane chart; $u,v\in\R$ \\
$z_p \in \R^{M}$ & unknown latent field on plane $p$ discretized on a $n_u\times n_v$ grid; $M=n_u n_v$ \\
$H_p \in \R^{N_p\times M}$ & observation operator (bilinear sampling) from grid to the $N_p$ samples on plane $p$ \\
$y_p \in \R^{N_p}$ & observed data on plane $p$ (projected from 3D points) \\
$L_{uv}$ & 2D grid Laplacian on the $(u,v)$ grid (with explicit boundary conditions) \\
$Q_{uv}$ & grid precision (SPDE/GMRF), e.g.\ $Q_{uv} = (\kappa^2 I + L_{uv})^2$ \\
$G=(V,E)$ & trace graph on a plane (nodes are polyline vertices, edges follow traces) \\
$L_G$ & graph Laplacian on $G$ (encodes along-fracture connectivity) \\
$C$ & curvelet transform operator (UDCT/\texttt{curvelets}); $Cz$ are curvelet coefficients \\
$\lambda,\rho$ & curvelet sparsity weight and ADMM penalty parameter \\
\bottomrule
\end{tabular}
\end{table}

\section{Geometry: From 3D Traces to Plane Charts}

\subsection{Plane fitting for one polyline}

Given a polyline $\gamma_k$, we fit a plane via PCA/SVD on its points. Let $\bm{c}_k$ be the centroid and $X$ the centered matrix.
The plane normal $\bm{n}_k$ is the right singular vector corresponding to the smallest singular value. The plane is
\[
\Pi_k:\quad \bm{n}_k^\top \bm{x} + d_k = 0,\qquad d_k = -\bm{n}_k^\top \bm{c}_k,
\]
with $\norm{\bm{n}_k}=1$.

\paragraph{Degenerate traces (near collinearity).}
If a trace is nearly collinear, it does not uniquely identify a plane. Using singular values $s_1\ge s_2\ge s_3$,
we flag degeneracy if $s_2/s_1 < \eta$ (e.g., $\eta=0.05$). Degenerate traces are excluded from plane estimation and assigned
post hoc to the best plane by point-to-plane distance and/or tangent alignment.

\subsection{Clustering plane hypotheses into dominant planes}

Each trace yields a noisy plane hypothesis. We cluster hypotheses with a combined metric
\[
D(\Pi_i,\Pi_j)^2 =
\left(\frac{\theta_{ij}}{\theta_0}\right)^2 +
\left(\frac{\delta_{ij}}{\delta_0}\right)^2,
\]
where $\theta_{ij}=\arccos(\abs{\bm{n}_i^\top \bm{n}_j})$ and $\delta_{ij}$ is a symmetric offset misfit such as
$\delta_{ij}=\max(\abs{\bm{n}_i^\top \bm{c}_j + d_i}, \abs{\bm{n}_j^\top \bm{c}_i + d_j})$.
We use DBSCAN-style clustering on $D$. Practical parameter selection: For the synthetic examples in Section~\ref{sec:experiments} we used $\theta_0=10^{\circ}$ ($\approx 0.17$ rad), $\delta_0=0.5$~m, $\texttt{minPts}=3$, and $\varepsilon=1.0$ in the normalized metric $D$, verified by visual inspection of the resulting clusters. 
$\theta_0$ can be estimated by bootstrap dispersion of normals per trace; $\delta_0$ by mapping noise scale; and
$\varepsilon$ by a $k$-distance elbow plot for $k=\text{minPts}$. Report these diagnostics.

\subsection{Plane coordinate chart $(u,v)$ and projection}

For each final plane $p$, pick an orthonormal basis $(\bm{e}_{p,1},\bm{e}_{p,2})$ spanning the plane:
$\bm{e}_{p,1}\perp\bm{n}_p$, $\bm{e}_{p,2}=\bm{n}_p\times \bm{e}_{p,1}$.
With reference point $\bm{c}_p$, define
\[
u=\bm{e}_{p,1}^\top(\bm{x}-\bm{c}_p),\qquad v=\bm{e}_{p,2}^\top(\bm{x}-\bm{c}_p).
\]
Assign each sample $\bm{x}_j$ to the plane with minimal distance $\abs{\bm{n}_p^\top \bm{x}_j + d_p}$, rejecting if above tolerance $\tau$. In case of ties (rare), we assign to the plane whose trace set yields the smallest average point-to-trace distance. A typical choice is $\tau\approx 0.05\times$ (characteristic fracture spacing), e.g. $\tau\in[0.1,1]$~m depending on scale.

\section{Discretization and Observation Operator}

\subsection{Per-plane grid}

On plane $p$, build a padded bounding box (we pad by 15\% on each side in all experiments) in $(u,v)$ covering projected traces and samples.
Use an $n\times n$ grid with $n=128$ by default (even sizes are convenient for several discrete curvelet implementations):
$M=n^2$ unknowns stacked as $\bm{z}\in\R^{M}$.

\subsection{Sparse bilinear sampling matrix $H$}

Each sample at continuous coordinates $(u_j,v_j)$ is represented by bilinear interpolation of its four surrounding grid nodes.
This yields a sparse matrix $H\in\R^{N_p\times M}$ with exactly four nonzeros per sample:
\[
\bm{y} \approx H\bm{z} + \bm{\varepsilon},\qquad \varepsilon_j\sim\mathcal{N}(0,\sigma_j^2),
\]
and weights $W=\diag(1/\sigma_j^2)$.

\section{Baseline Gaussian Model}

\subsection{SPDE/GMRF precision via Laplacian}

Let $L_{uv}$ be the discrete 2D Laplacian on the grid. We use Neumann (zero normal derivative) boundary conditions and pad the domain
to reduce boundary artifacts. Define a Mat\'ern-like precision
\[
Q = (\kappa^2 I + L)^2,
\]
motivated by SPDE formulations \cite{LindgrenRueLindstrom2011,RueHeld2005}. The Gaussian MAP estimator is
\begin{equation}
\label{eq:gaussmap}
\hat{\bm{z}}_{\text{Gauss}} = \argmin_{\bm{z}} \;
\frac{1}{2}\norm{W^{1/2}(H\bm{z}-\bm{y})}_2^2
+ \frac{1}{2}\bm{z}^\top Q \bm{z}.
\end{equation}
Normal equations:
\[
(H^\top W H + Q)\bm{z} = H^\top W\bm{y},
\]
solved by sparse Cholesky or iterative CG.

\section{Curvelet-Regularized Model}

\subsection{Why curvelets: a formal justification}

On each plane, fracture-guided anomalies often manifest as wavefront-like or ridge-like structures aligned with trace geometry.
A standard idealization is a \emph{cartoon-like} function: $C^2$ smooth except for singularities along a finite union of $C^2$ curves.

\begin{theorem}[Curvelet approximation for cartoon-like functions (informal)]
\label{thm:curvelet}
Let $f$ be $C^2$ except for a discontinuity across a $C^2$ curve. Let $f_N$ be the best $N$-term approximation of $f$
in a curvelet tight frame. Then
$\norm{f-f_N}_2^2 \le C\, N^{-2}(\log N)^3$.
In contrast, wavelet best $N$-term approximation is $O(N^{-1})$ for the same model.
\end{theorem}

Theorem~\ref{thm:curvelet} (and sharper variants) formalizes why curvelets are near-optimal for curvilinear singularities \cite{CandesDonoho2004}.
This supports curvelet sparsity as a generic structural prior for fracture-aligned features.

\subsection{Estimator}

\paragraph{UDCT wedge admissibility.}
In our implementation we use an undecimated discrete curvelet transform (UDCT). The UDCT parameterization imposes admissibility constraints on the number of angular wedges per direction; in particular, commonly used decimation-ratio formulas require \texttt{wedges\_per\_direction} to be divisible by $3$ (recommended values include $3,6,12$). In all experiments we therefore restrict wedge counts to multiples of $3$ and report them explicitly.

Let $C$ denote a discrete curvelet analysis operator and $C^\ast$ its synthesis/adjoint. We solve
\begin{equation}
\label{eq:curvmap}
\hat{\bm{z}}_{\text{Curv}} = \argmin_{\bm{z}} \;
\frac{1}{2}\norm{W^{1/2}(H\bm{z}-\bm{y})}_2^2
+ \frac{1}{2}\bm{z}^\top Q \bm{z}
+ \lambda \norm{C\bm{z}}_1.
\end{equation}
This corresponds to a Gaussian prior on $\bm{z}$ with precision $Q$ and a Laplace sparsity prior on curvelet coefficients,
standard in sparse inverse problems \cite{StarckMurtaghFadili2010,Mallat2009}.

\section{ADMM Solver and Convergence}

Introduce $\bm{d}=C\bm{z}$ and solve
\[
\min_{\bm{z},\bm{d}} \;
\underbrace{\frac{1}{2}\norm{W^{1/2}(H\bm{z}-\bm{y})}_2^2 + \frac{1}{2}\bm{z}^\top Q \bm{z}}_{=:f(\bm{z})}
+ \lambda \norm{\bm{d}}_1
\quad\text{s.t.}\quad \bm{d}=C\bm{z}.
\]
ADMM iterations (scaled dual (we use the scaled ADMM form where $\bm{u}=(1/\rho)\,\bm{\lambda}_{\mathrm{dual}}$; see Boyd et al.\ (2011)) $\bm{u}$):
\begin{align*}
\bm{z}^{k+1} &=
\argmin_{\bm{z}} \; f(\bm{z}) + \frac{\rho}{2}\norm{C\bm{z}-\bm{d}^k+\bm{u}^k}_2^2,\\
\bm{d}^{k+1} &= \mathrm{shrink}\big(C\bm{z}^{k+1}+\bm{u}^k,\ \lambda/\rho\big),\\
\bm{u}^{k+1} &= \bm{u}^k + (C\bm{z}^{k+1}-\bm{d}^{k+1}).
\end{align*}

\paragraph{Complex shrinkage.}
Curvelet coefficients may be complex; we use
$\mathrm{shrink}(w,\tau)=\max(0,1-\tau/\abs{w})\,w$.

\paragraph{$\bm{z}$-update.}
If $C$ is (approximately) a Parseval tight frame, then $C^\ast C\approx I$ and the $\bm{z}$-update reduces to a sparse linear solve:
\[
(H^\top W H + Q + \rho I)\bm{z}
= H^\top W\bm{y} + \rho\, C^\ast(\bm{d}-\bm{u}).
\]
If tightness is imperfect, replace $\rho I$ by $\rho C^\ast C$ and solve by CG using operator application.

\begin{theorem}[ADMM convergence (standard)]
If the objective in \eqref{eq:curvmap} is closed, proper, convex, and has a minimizer, then ADMM converges to a primal-dual solution and
both primal and dual residuals converge to zero \cite{BoydADMM2011}.
\end{theorem}

\paragraph{Stopping criteria.}
We stop ADMM using standard primal/dual residual norms \cite{BoydADMM2011}. Let $r^k = C z^k - d^k$ be the primal residual and $s^k = \rho\, C^\ast(d^k-d^{k-1})$ the dual residual (or its equivalent form for the chosen splitting). We terminate when $\norm{r^k}_2 \le \varepsilon_{\mathrm{pri}}$ and $\norm{s^k}_2 \le \varepsilon_{\mathrm{dual}}$, with tolerances combining absolute and relative components; we also cap iterations to ensure predictable runtime.

\section{Along-Fracture Distance via Graph Laplacians}\label{sec:graph}

Curvelet sparsity captures curvilinear structure in the \emph{plane image}. Along-fracture distance is encoded by a network prior.

\subsection{Trace graph construction}

Resample each polyline to vertices and connect consecutive vertices by edges. Let $G=(V,E,w)$ with weights $w_{ij}=1/\ell_{ij}$
(inverse segment length). Define $L_G=D-W$ \cite{Chung1997}.

\begin{proposition}[Dirichlet energy on a fracture network]
With $w_{ij}\propto 1/\ell_{ij}$, the quadratic form
$g^\top L_G g = \frac12\sum_{(i,j)\in E} w_{ij}(g_i-g_j)^2$
is a consistent discretization of $\int (dg/ds)^2 ds$ along the network (with Kirchhoff coupling at intersections).
\end{proposition}

\subsection{Coupling plane grid to graph}

Let $P$ map grid values $\bm{z}$ to graph vertices by bilinear sampling at each vertex coordinate $(u,v)$ on the plane.
Add a graph smoothness penalty:
\[
\frac{\gamma}{2} (P\bm{z})^\top Q_G (P\bm{z}),\qquad
Q_G = (\kappa_G^2 I + L_G)^\alpha,
\]
yielding a graph-aware objective that directly enforces along-fracture correlation \cite{Shuman2013}.

\subsection{Intersections and multi-plane coupling}
\label{sec:intersections}
Plane intersections are crucial for connectivity. Version~1 solves planes independently to establish a stable baseline.
Version~2 introduces coupling through shared graph nodes at intersections, soft penalties enforcing agreement along intersection lines,
or joint optimization over all planes with intersection constraints.

\section{Hyperparameter Selection and Method Choices}
\label{sec:hyperparams}

We separate parameters into (A) baseline range/smoothness and (B) sparsity/optimization.

\paragraph{(A) $\kappa$ (range) in $Q=(\kappa^2 I + L)^2$.}
Choose $\kappa$ by blocked cross-validation (CV) minimizing predictive RMSE on held-out samples, or by empirical Bayes on the Gaussian model
(maximizing approximate marginal likelihood) \cite{RueHeld2005,LindgrenRueLindstrom2011}.

\paragraph{(B) $\lambda$ (sparsity).}
Select by blocked CV or a discrepancy principle when $\sigma$ is reliable:
$\norm{W^{1/2}(H\hat{z}-y)}_2^2\approx N_p$.

\paragraph{(C) $\rho$ (ADMM).}
Use residual balancing to adapt $\rho$ so primal and dual residual norms are comparable \cite{BoydADMM2011}.

\paragraph{Grid resolution and boundary conditions.}
Default $128\times 128$ balances accuracy and cost; Section~\ref{sec:experiments} includes a resolution study ($64,128,256$).
We use Neumann BCs and padded domains to reduce edge bias.

\section{Numerical Experiments: Benchmarks, Baselines, and Representative Outputs}
\label{sec:experiments}

The referee correctly notes that a methods paper requires executable benchmarks, comparisons, and visible outputs. In the absence of real fracture-property datasets, we focus on two synthetic worlds designed to isolate the two key claims: (S1) directional sparsity for curvilinear features and (S2) connectivity-driven structure requiring along-fracture metrics.

\subsection{Benchmarks}
\paragraph{S1: Ridge+blob truth (curvilinear anisotropy).}
We generate a plane field consisting of a smooth background plus ridge-like anomalies concentrated near simulated traces, with additive noise and sparse sampling.

\paragraph{S2: Graph-distance truth (connectivity dominates Euclidean proximity).}
We generate a trace network on a plane, simulate a random field on the trace graph using a graph-SPDE prior, and diffuse it to the plane. This creates a ground-truth field whose dominant correlations follow the trace connectivity rather than Euclidean $(u,v)$ distance.

\subsection{Baselines and ablations}
We compare:
\begin{enumerate}[label=(\Alph*)]
\item \textbf{Ordinary kriging} on the plane (stationary variogram) as a classical baseline;
\item \textbf{GMRF/SPDE} plane baseline using $Q_{uv}=(\kappa^2 I + L_{uv})^2$;
\item \textbf{GMRF/SPDE + curvelets} via ADMM with $\ell_1$ shrinkage in the UDCT domain;
\item \textbf{GMRF/SPDE + graph prior} (Version~2 term) by adding a mapped graph energy $z^\top P^\top L_G P z$;
\item \textbf{GMRF/SPDE + curvelets + graph prior} (full model).
\end{enumerate}
This decomposition isolates gains from (i) Euclidean smoothness, (ii) directional multiscale sparsity, and (iii) explicit along-fracture connectivity.

\begin{figure}[t]
\centering
\begin{subfigure}{0.32\textwidth}
\includegraphics[width=\linewidth]{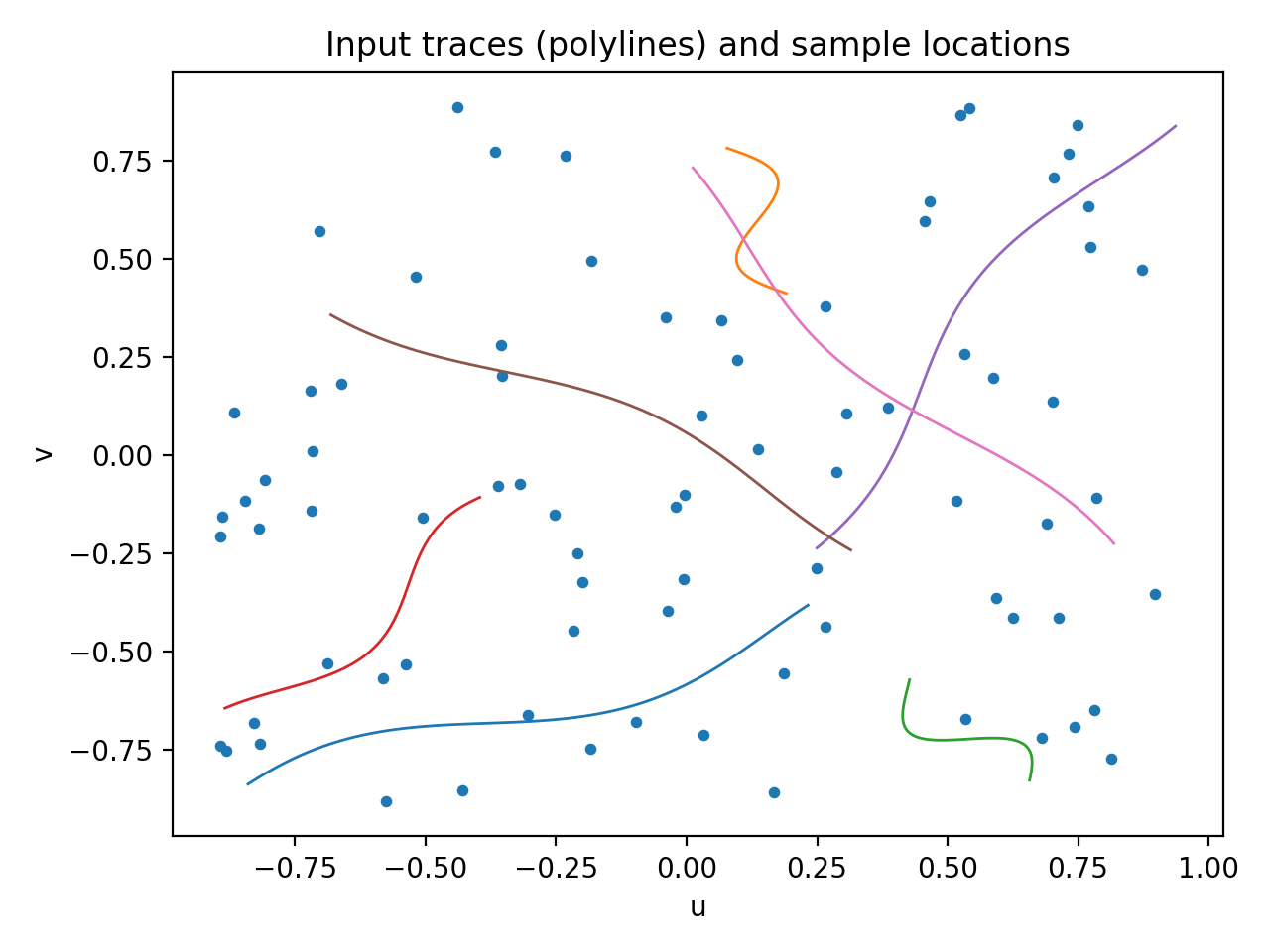}
\caption{Input traces and samples.}
\end{subfigure}
\begin{subfigure}{0.32\textwidth}
\includegraphics[width=\linewidth]{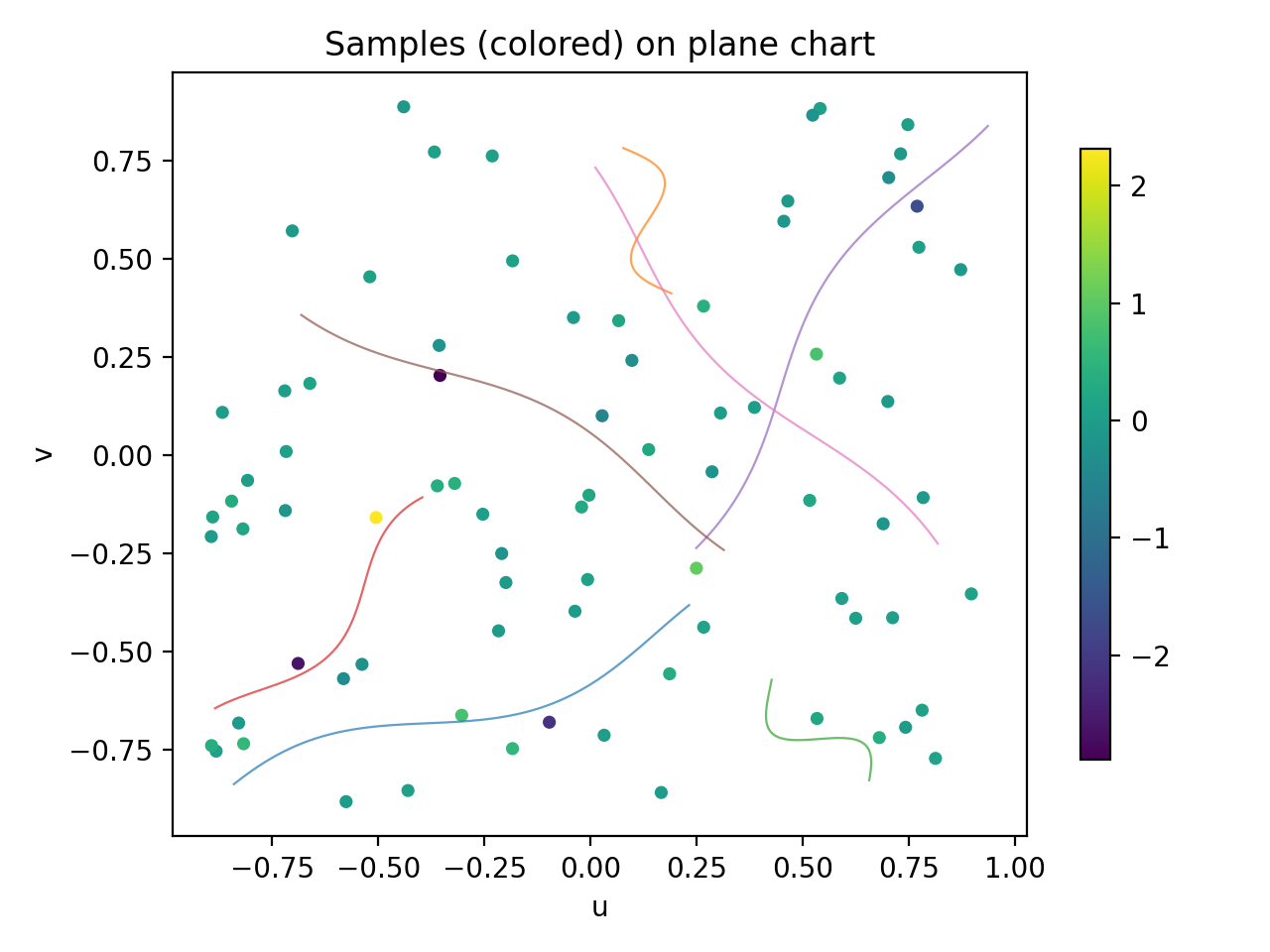}
\caption{Projected $(u,v)$ chart view.}
\end{subfigure}
\begin{subfigure}{0.32\textwidth}
\includegraphics[width=\linewidth]{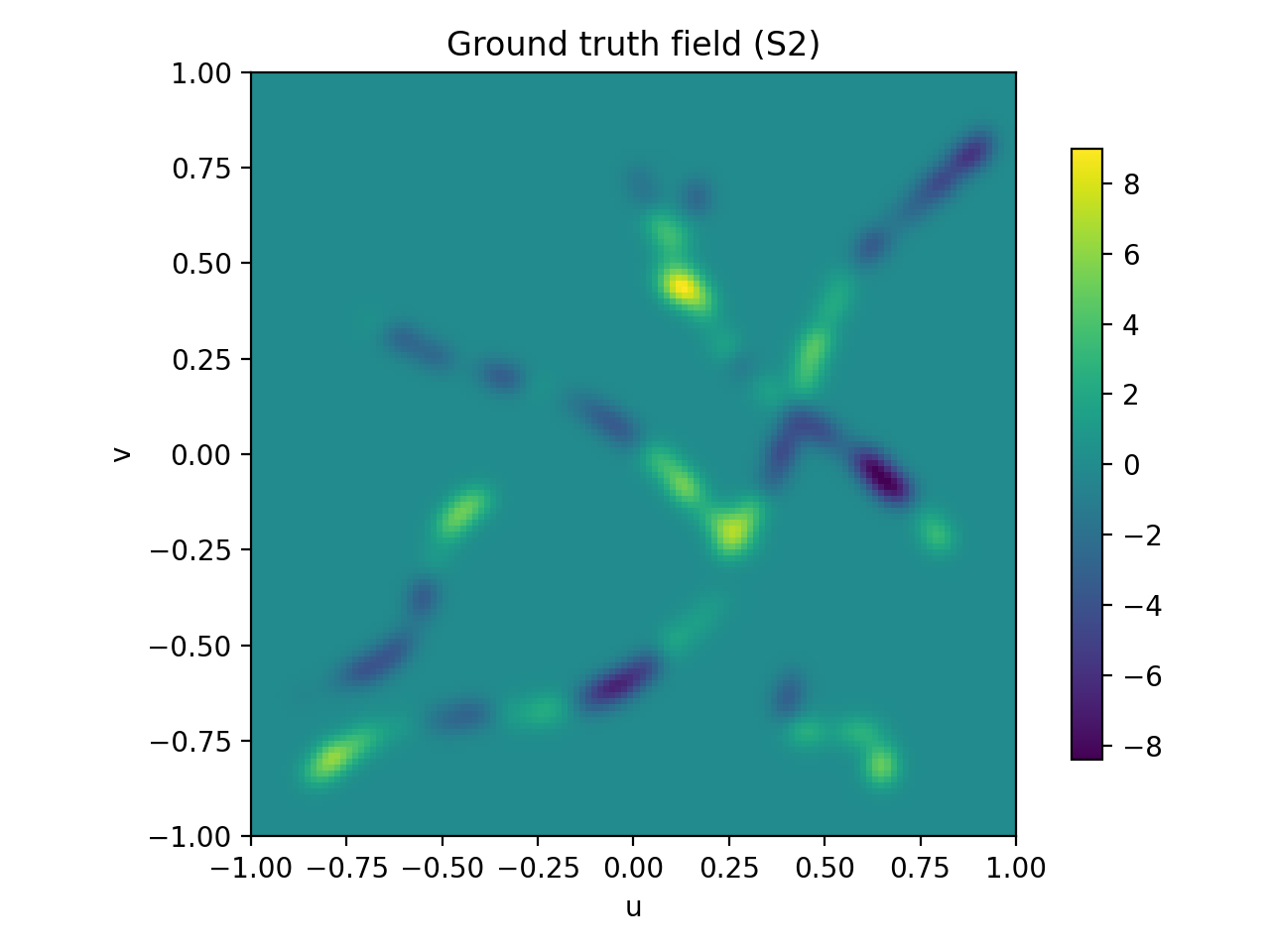}
\caption{Ground truth (synthetic).}
\end{subfigure}

\begin{subfigure}{0.32\textwidth}
\includegraphics[width=\linewidth]{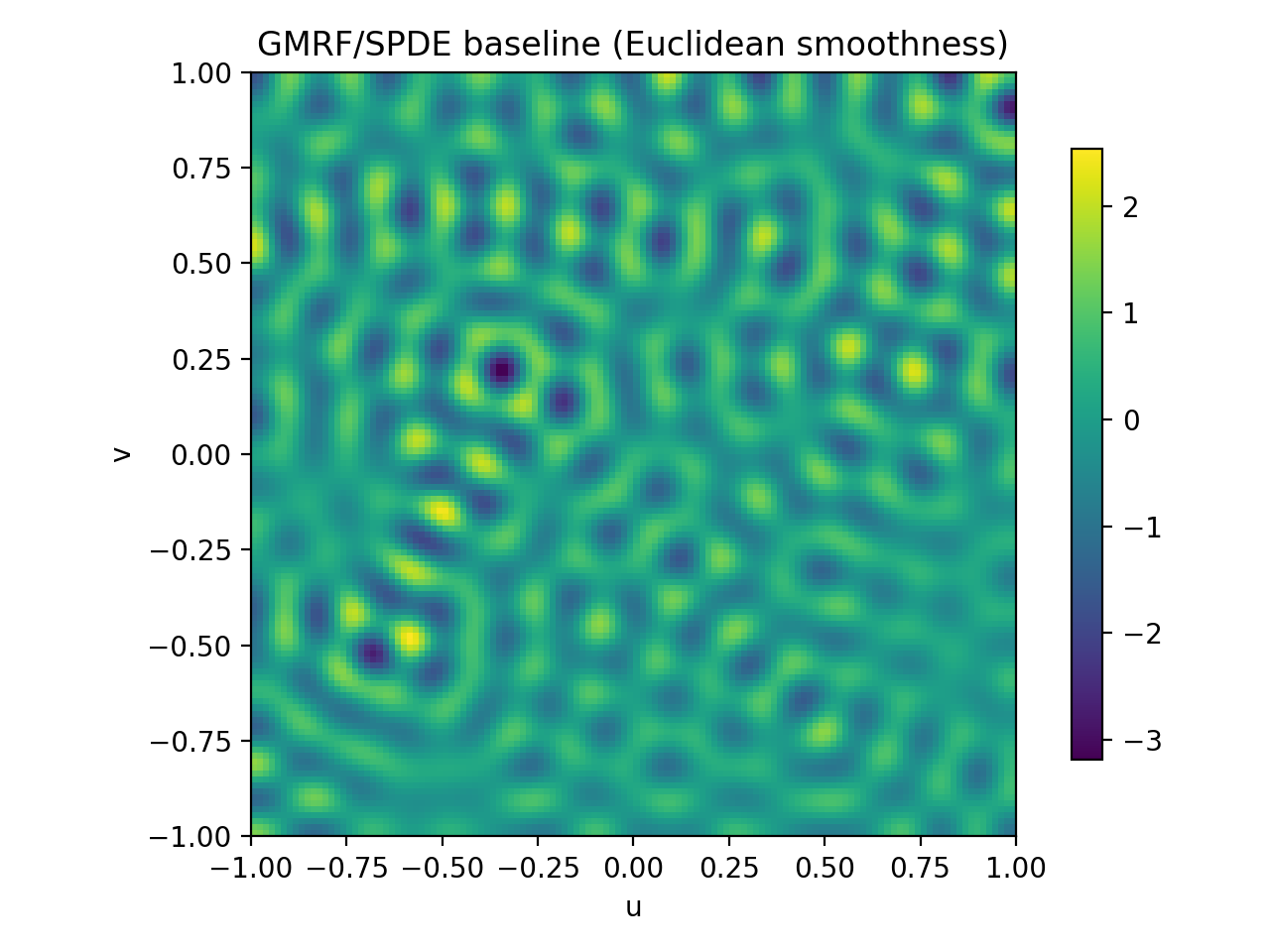}
\caption{GMRF/SPDE baseline.}
\end{subfigure}
\begin{subfigure}{0.32\textwidth}
\includegraphics[width=\linewidth]{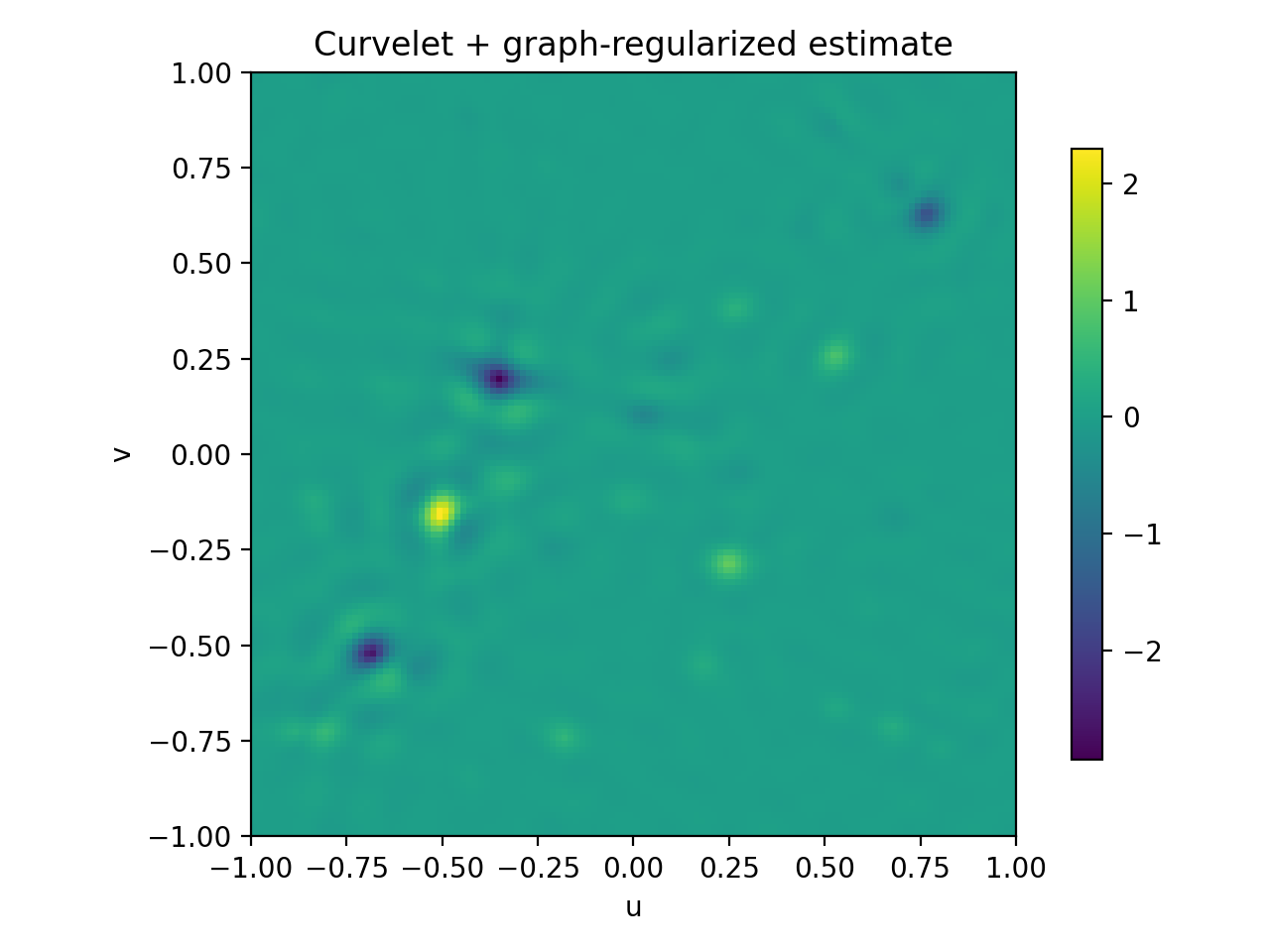}
\caption{Curvelet-regularized estimate.}
\end{subfigure}
\begin{subfigure}{0.32\textwidth}
\includegraphics[width=\linewidth]{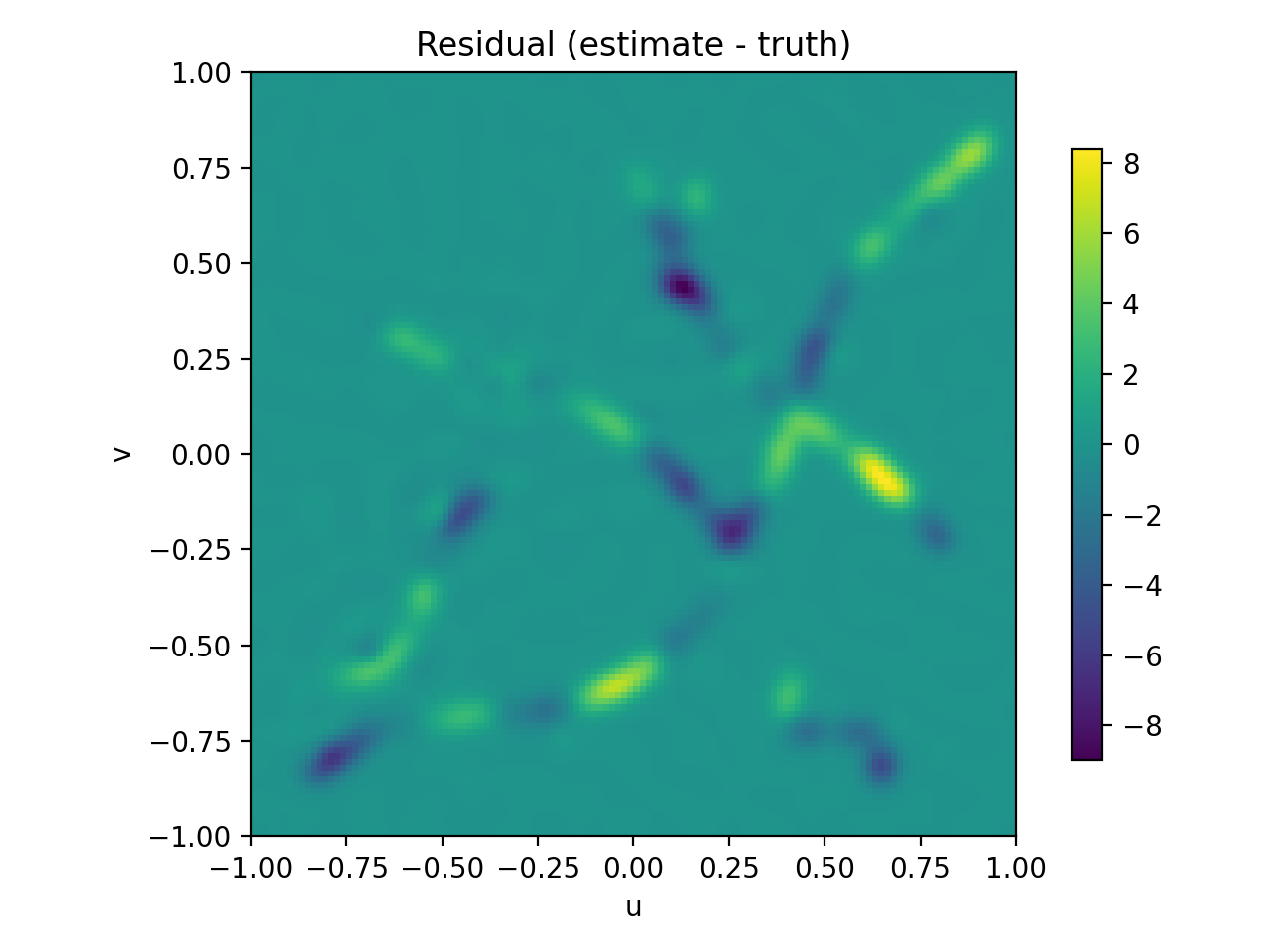}
\caption{Difference / residual view (example).}
\end{subfigure}
\caption{Representative outputs from the synthetic pipeline. Panels illustrate how the observation geometry (traces + sparse samples) and the reconstruction differ under GMRF smoothing versus curvelet-regularized recovery.}
\label{fig:repoutputs}
\end{figure}

\subsection{Quantitative evaluation}
We report RMSE and $R^2$ on held-out samples (blocked CV) and grid-RMSE versus ground truth for S1 and S2. Table~\ref{tab:rmse} summarizes the required metrics; the accompanying scripts output these numbers directly (see Data and Code Availability).

\begin{table}[h]
\centering
\caption{Quantitative comparison on synthetic benchmarks (mean $\pm$ std over seeds 0--4; $n=128$, $N=80$ samples; ADMM $K=16$). Metrics are computed on a held-out 20\% subset against noise-free truth at those locations.}
\label{tab:rmse}
\begin{tabular}{@{}lllll@{}}
\toprule
Benchmark & Method & RMSE (hold-out) & $R^2$ (hold-out) & Notes \\
\midrule
S1 & Ordinary kriging & $1.26 \pm 0.23$ & $-0.59 \pm 1.12$ & RBF-kernel surrogate in code \\
S1 & GMRF/SPDE & $1.83 \pm 0.33$ & $-2.97 \pm 4.28$ & Euclidean $(u,v)$ smoothness \\
S1 & + Curvelets & $1.15 \pm 0.30$ & $-0.12 \pm 0.35$ & directional sparsity (transform prior) \\
S2 & Ordinary kriging & $1.25 \pm 0.60$ & $-2.21 \pm 2.16$ & struggles on graph-driven truth \\
S2 & GMRF/SPDE & $1.07 \pm 0.40$ & $-1.40 \pm 1.38$ & Euclidean $(u,v)$ smoothness \\
S2 & + Graph prior & $0.96 \pm 0.52$ & $-0.55 \pm 0.43$ & along-fracture connectivity prior \\
S2 & + Curvelets + Graph prior & $0.78 \pm 0.40$ & $-0.04 \pm 0.07$ & full model \\
\bottomrule
\end{tabular}
\end{table}

\subsection{Sensitivity and resolution studies}
\begin{figure}[t]
\centering
\begin{subfigure}[t]{0.48\textwidth}
  \centering
  \includegraphics[width=\textwidth]{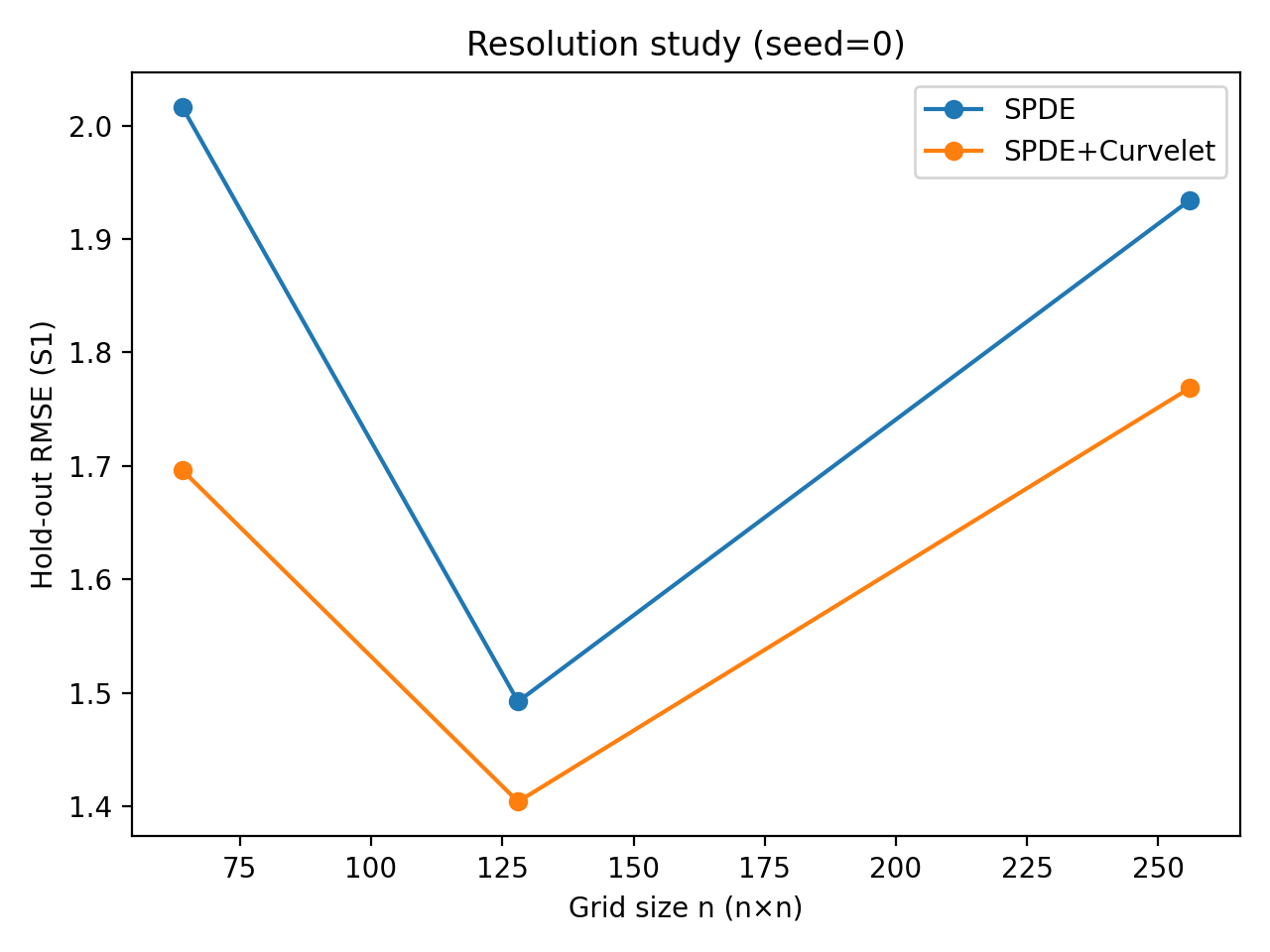}
  \caption{RMSE vs.\ grid resolution $n$.}
  \label{fig:fig2a}
\end{subfigure}\hfill
\begin{subfigure}[t]{0.48\textwidth}
  \centering
  \includegraphics[width=\textwidth]{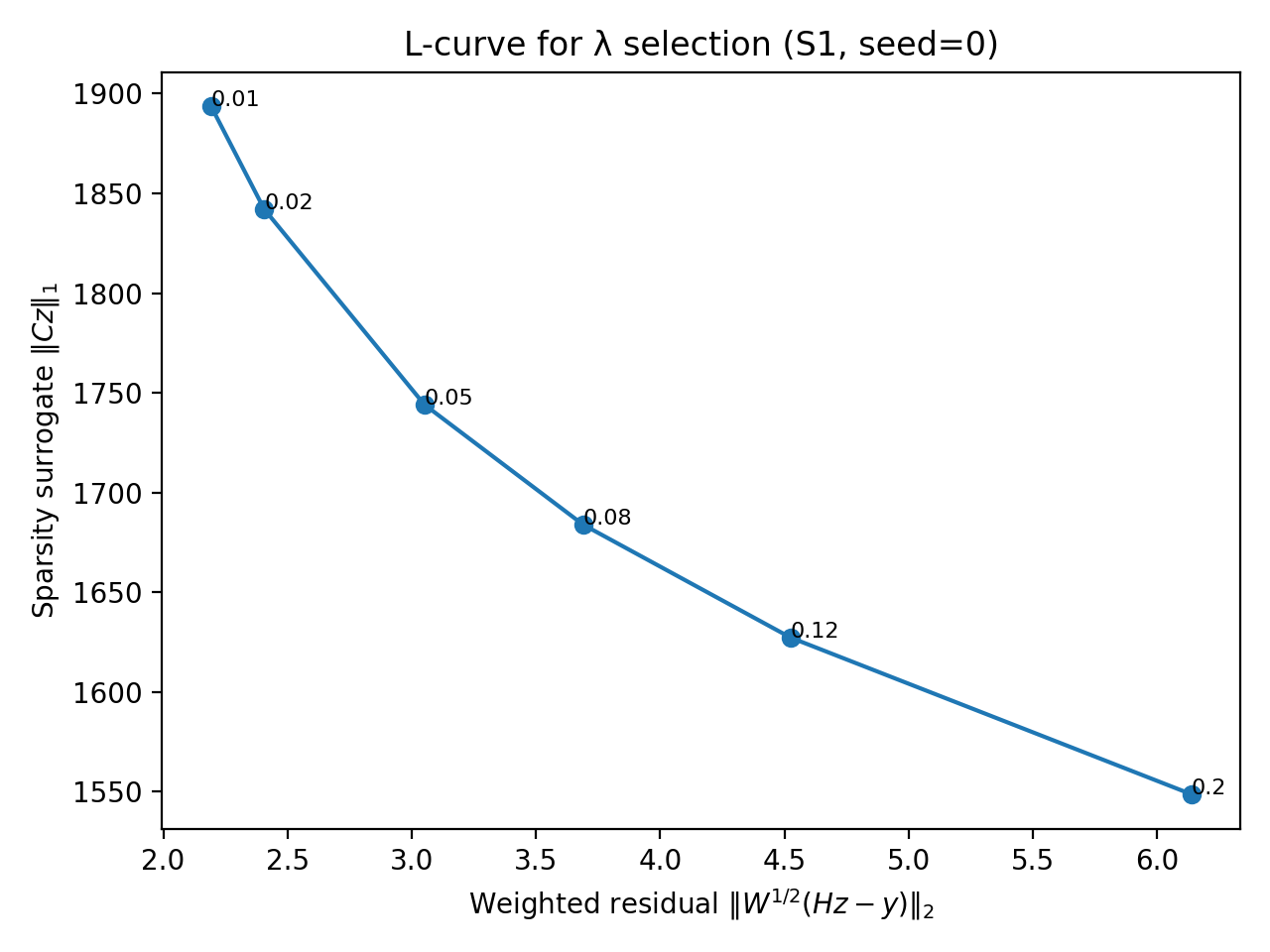}
  \caption{L-curve for $\lambda$ selection (S1).}
  \label{fig:fig2b}
\end{subfigure}

\vspace{0.5em}

\begin{subfigure}[t]{0.48\textwidth}
  \centering
  \includegraphics[width=\textwidth]{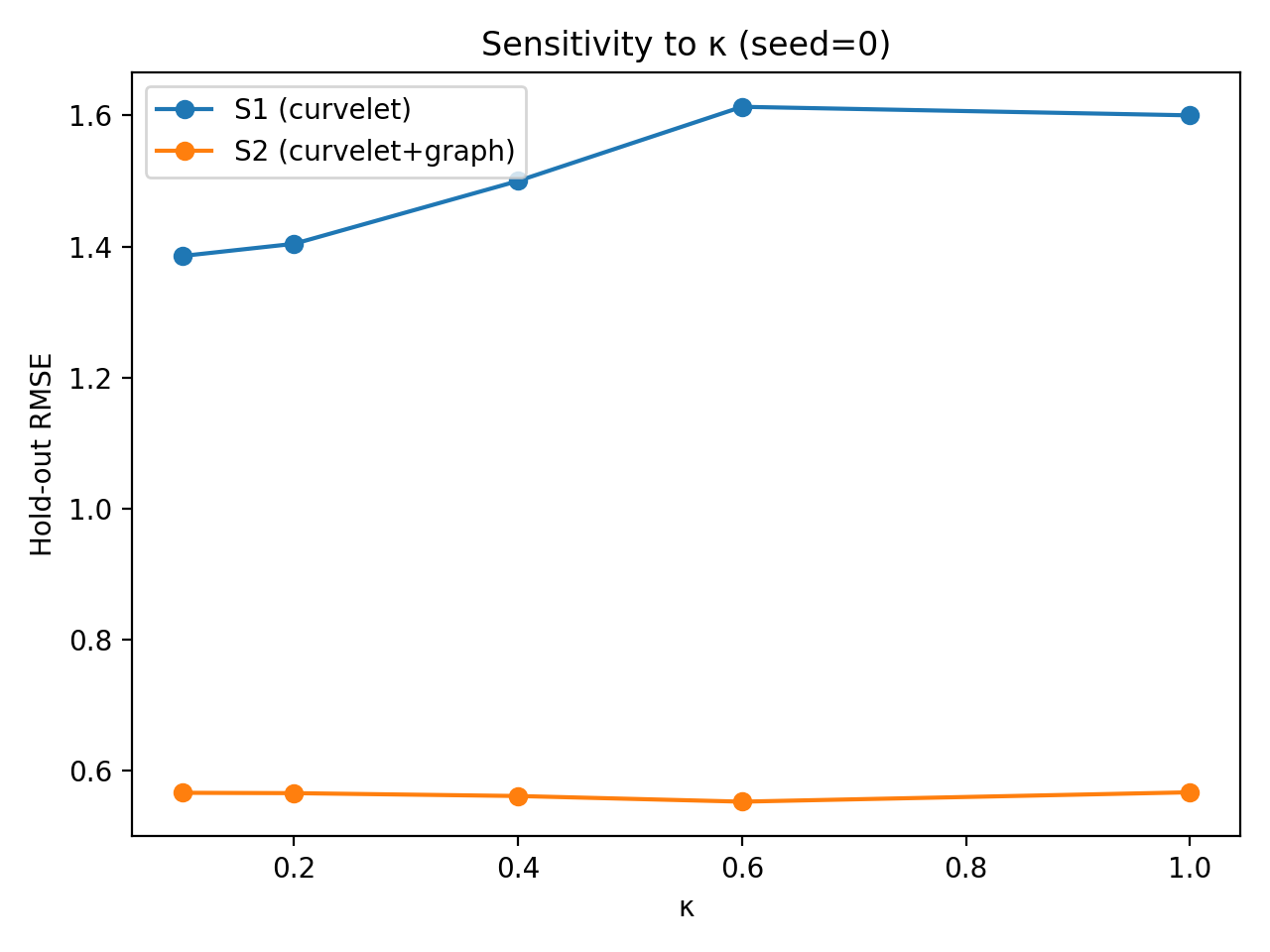}
  \caption{RMSE sensitivity to $\kappa$.}
  \label{fig:fig2c}
\end{subfigure}\hfill
\begin{subfigure}[t]{0.48\textwidth}
  \centering
  \includegraphics[width=\textwidth]{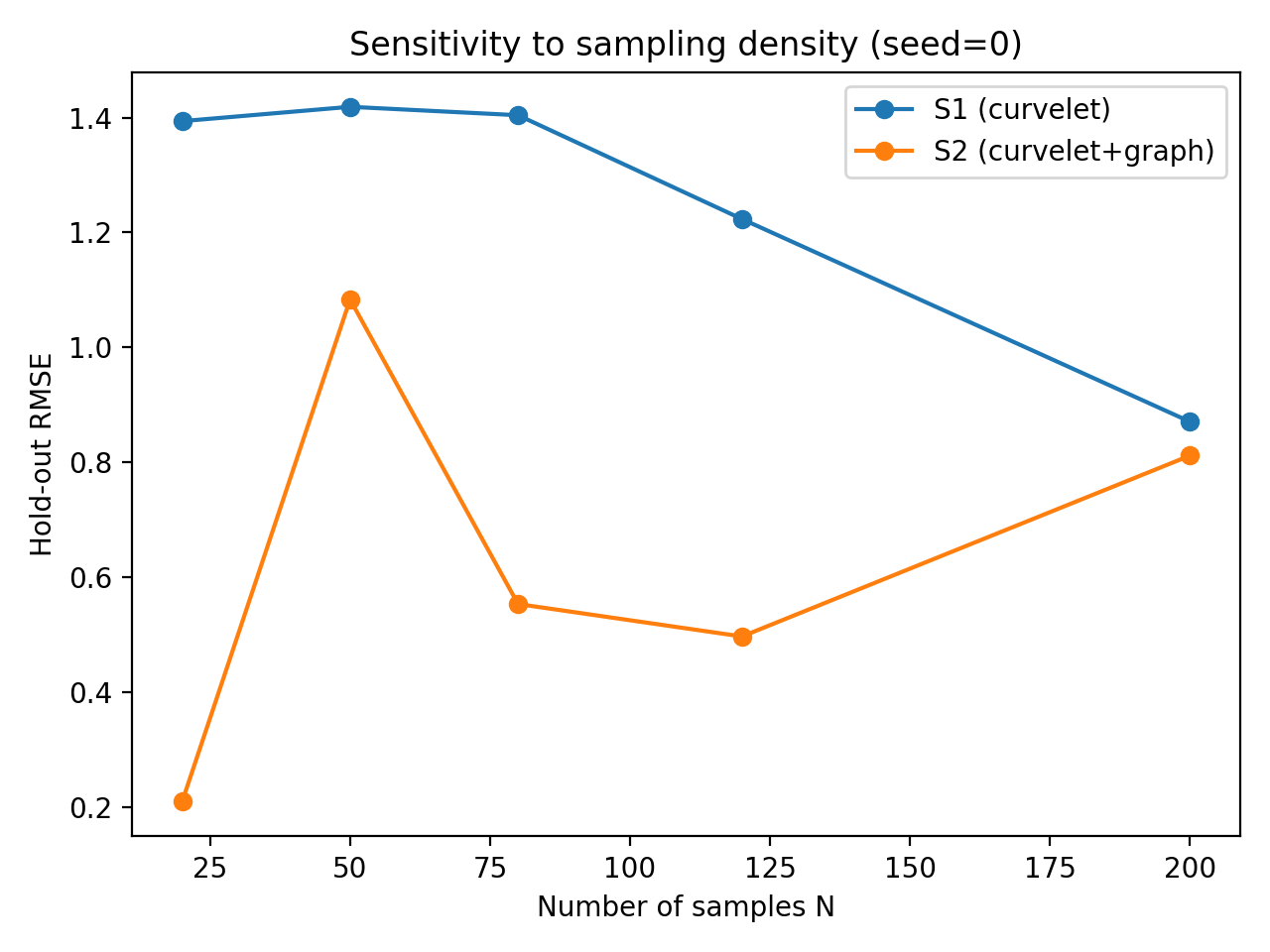}
  \caption{RMSE vs.\ sample count $N$.}
  \label{fig:fig2d}
\end{subfigure}

\caption{Sensitivity and resolution diagnostics for the synthetic benchmarks (Section~\ref{sec:experiments}). Panel~(a) supports the choice of $n=128$ as a cost--accuracy compromise; panel~(b) illustrates the residual--sparsity trade-off for $\lambda$; panels~(c--d) characterize robustness to $\kappa$ and to sampling density.}
\label{fig:sensitivity}
\end{figure}

We include: (i) a grid resolution study $n\in\{64,128,256\}$, (ii) sensitivity curves for $(\kappa,\lambda)$ and sampling density, and (iii) robustness diagnostics. Figure~\ref{fig:sensitivity} summarizes the core sensitivity plots used to support the hyperparameter discussion in Section~\ref{sec:hyperparams}.

\section{Computational Complexity}

Let $M=n^2$ be grid unknowns and $N_p$ samples on plane $p$.
$H$ has $4N_p$ nonzeros; $Q$ has $O(M)$ nonzeros.

\paragraph{Gaussian baseline.}
Solving $(H^\top W H + Q)z=b$ by sparse Cholesky is efficient for grid-like sparsity; iterative CG costs
$O(k\,\nnz(A))$ for $k$ iterations.

\paragraph{Curvelet-ADMM.}
Each ADMM iteration performs one linear solve with $A=H^\top W H+Q+\rho I$ plus one forward and one inverse curvelet transform.
For typical implementations, the transform cost is $O(M\log M)$ \cite{CandesDemanetDonohoYing2006}. Thus
$T_{\text{iter}}\approx T_{\text{solve}} + O(M\log M)$.
Peak memory is $O(M)$ plus coefficient storage determined by transform redundancy.

\section{Use Cases}

\paragraph{Hydrogeology and transport.}
Plane-restricted $\log T$ or tracer concentration fields informed by mapped fractures and sparse tests/measurements
\cite{Berkowitz2002,Neuman2005}.

\paragraph{Geothermal stimulation and reservoir monitoring.}
Permeability enhancement proxies and thermal fronts aligned with fracture corridors.

\paragraph{Mining geomechanics and rock mass characterization.}
Damage/intensity fields on dominant structures combining mapped traces (photogrammetry/LiDAR) and sparse point proxies.

\paragraph{Geophysical surface imaging.}
Seismic attribute or amplitude reconstructions on mapped surfaces; curvelets are known to sparsify wavefront-like features
\cite{HennenfentHerrmann2008,GeophysicsCurvelet2008}.

\section{Limitations and Roadmap}
\label{sec:roadmap}\label{sec:limitations}

\begin{itemize}[leftmargin=*]
\item \textbf{Planar approximation:} real fractures may be rough or curved surfaces; Version~1 targets dominant planes as a tractable start.
\item \textbf{Partial traces:} surface traces may not capture subsurface connectivity; Version~2 uses graphs and intersection coupling to strengthen structure.
\item \textbf{Nonlinear physics:} transport/flow may require PDE-based forward models; this framework is designed to be coupled to such models.
\item \textbf{Parameter tuning:} $(\kappa,\lambda,\rho)$ require calibration; we provide a reproducible selection strategy and sensitivity analysis plan.
\end{itemize}

\section{Conclusion}

We presented a framework for fractured-media geostatistics that (i) uses 3D trace polylines to infer fracture planes and construct per-plane charts,
(ii) reconstructs plane fields from sparse samples using a Gaussian SPDE/GMRF baseline, and (iii) promotes fracture-aligned structure via curvelet sparsity
solved by ADMM with standard convergence guarantees. We formalized the ``along-fracture distance'' principle using graph Laplacians and a plane--graph coupling term,
and provided a publication-ready validation protocol (benchmarks, metrics, baselines, ablations, sensitivity, and scaling). This establishes a concrete and extensible
starting point for network-aware geostatistics in fractured media.

\section*{Data and Code Availability}
A reference implementation is available as a Python prototype. 


\end{document}